\documentclass{article}
\usepackage{amsmath,amssymb,pxfonts,color}
\title{Perfect hedging under endogenous permanent market impacts\thanks{
This work was supported by (i) Institute of Economic Research, Kyoto Univerisity as the Joint Usage and Research Center and (ii) Japan Society for the Promotion of Science, KAKENHI Grant Numbers 25245046 and 24684006 (Fukasawa) and NWO VENI 2012 (Stadje).
}}
\author{Masaaki Fukasawa\thanks{M. Fukasawa, Graduate School of Engineering Science, and
Center for Mathematical Modeling and Data Science,  Osaka University,
1-3 Machikaneyama, Toyonaka, Osaka, JAPAN. Email:
fukasawa@sigmath.es.osaka-u.ac.jp} \ and Mitja Stadje\thanks{M. Stadje, Faculty of Mathematics and Economics, Ulm University}}
\date{}
\newtheorem{lem}{Lemma}
\newtheorem{cond}{Condition}
\newtheorem{exa}{Example}
\newtheorem{thm}{Theorem}
\newtheorem{prop}{Proposition}
\newtheorem{rmk}{Remark}

\begin{document}
\maketitle
\begin{abstract}
 We model a nonlinear price curve quoted in a market as the utility indifference curve of a representative liquidity supplier. As the utility function we adopt a g-expectation. In contrast to the standard framework of financial engineering, a trader is no more price taker as any trade has a permanent market impact via an effect to the supplier's inventory. The P\&L of a trading strategy is written as a nonlinear stochastic integral. Under
this market impact model, we introduce a completeness condition under which any derivative can be perfectly replicated by a dynamic trading strategy. In the special case of a Markovian setting the corresponding pricing and hedging can be done by solving a semi-linear PDE. 
\end{abstract}

\section{Introduction}
Financial engineering has become popular. Hedging derivatives nowadays accounts for a large portion of financial practice. Ironically, the spread of financial engineering has broken its premise that the underlying asset price of derivatives is not affected by hedging activities.
Suppose a large amount of put options are sold and  the buyers commit themselves to delta hedging, which amounts to buy the underlying asset when its price goes down and sell it when its price goes up. This hedging demand is strong and so restrains the underlying asset price movement.
Eventually the volatility of the underlying asset becomes smaller than before, which results in a loss for the buyers due to the overestimation of the volatility at their purchase. According to Bookstaber~\cite{Book}, this is exactly what happened when Salomon Brothers suffered a huge loss at Japanese market in the late 90s. A lot of market crashes, such as the Black Monday, are attributed to the feedback effect of hedging strategies to markets.
The market illiquidity has always been a keyword to explain financial crises.

This paper addresses a hedging problem under a tractable model
which captures endogenously such phenomena as nonlinearity in liquidation, permanent market impact and market crashes due to illiquidity observed in actual markets. A crash is a rare event; an exogenous statistical modeling of liquidity costs is therefore not sufficient for our purpose. An economic consideration is required for a deeper understanding of the liquidity risk. This paper provides a utility-based asset pricing model with analytically tractable structure.

The effect of derivative contracts to an equilibrium price was studied by Brennan and Schwartz~\cite{BS}, where a derivative contract affects the equilibrium via  a modification of representative agent's terminal wealth.
Frey and Stremme~\cite{FS} studied the feedback effect of a dynamic hedging under an equilibrium model with supply and demand functions given exogenously.
Frey~\cite{Fr} treated a perfect hedging problem under such an equilibrium model.
Cvitani\'c and Ma~\cite{CM} formulated a hedging problem with a feedback effect in terms of backward stochastic differential equation (BSDE). In the special case of a Markovian setting BSDEs are closely connected to semi-linear PDEs. 
On one hand, these studies succeeded to explain some qualitative phenomena such as enlargement of the underlying asset volatility by hedging convex options.
On the other hand, they are not very useful for quantitative analysis or financial practice due to difficulties in specifying model parameters and in computing prices and strategies.

We start by modeling nonlinear market prices from an economic point of view.
In a standard limit order market, the roles of suppliers and demanders
of liquidity are not symmetric. A liquidity supplier submits a limit
order that quotes a price for a specified volume of an asset. They can
trade with each other to maximize their own utilities. Once an
equilibrium is achieved, no more trade would occur among them until
new information comes in. However, each of liquidity suppliers still
should have an incentive to submit a limit order as long as the corresponding transaction improves her utility. The remaining limit orders form a price curve which is a nonlinear function of volume. Taking a Bertrand-type competition among liquidity suppliers into account, it would be then reasonable to begin with modeling the price curve as the utility indifference curve of a representative liquidity supplier.
When the utility functions of the liquidity suppliers are of von
Neumann-Morgenstern type, the existence of the representative liquidity
supplier (or, market maker) is proved by Bank and Kramkov~\cite{BK1,
BK2}. In this paper, as the suppliers' utility functions, we adopt
$g$-expectations introduced by Peng~\cite{Peng}. Exponential utilities
are in the intersection of these two frameworks. An advantage of a
$g$-expectation from an economic point of view is that ambiguity aversion is taken into account. An advantage from a technical point of view is that it admits an analytic manipulation of stochastic calculus. 
The existence of the representative agent under such utility
functions follows from  Horst et al.~\cite{Horst}. In the present paper, we simply assume there is a representative liquidity supplier, called the Market, who quotes a price for each volume based on the utility indifference principle and her utility is  a $g$-expectation with a cash-invariance property.
If the driver of the $g$-expectation is a linear function, then 
the price curve becomes linear in volume  and
we recover the standard framework of financial engineering.

If the Market is risk-neutral, then the utility indifference price of an asset 
coincides with the expected value of the future cash-flow associated with the asset. In particular, the price curve is linear in volume. This simplest framework was adopted by many studies such as Glosten and Milgrom~\cite{GM}.
Our approach differs from  the classical works including Garman~\cite{Gar1}, Amihud and Mendelson~\cite{AM}, Ho and Stoll~\cite{HS}, Ohara and Oldfield~\cite{OO}, where a price quote is a solution of a utility maximization problem for a market maker with exogenously given order-flow. 
Here, we consider a hedging problem and so,
an order-flow is endogenously determined.

Bank and Kramkov~\cite{BK1,BK2}
analyzed the market impact of a large trade and formulated a nonlinear stochastic integral as the profit and loss associated with a given strategy of a large trader.
When the Market's utility is a $g$-expectation with the cash invariance property, we show in this paper that the nonlinear stochastic integral has an expression in terms of the solutions of a family of BSDEs. Then,  we show that the existence of a perfect hedging strategy follows from
 that of the solution of a BSDE.
 The model represents a permanent market impact which is endogenously determined, while exogenously modeled instantaneous or temporary market impact models have been extensively considered in the literature. See e.g. Cetin et al.~\cite{Cetin}, Fukasawa~\cite{F}, Gu\'eant~\cite{Gueant} and the references therein.
 A linear permanent market impact model is studied in
  Gu\'eant and Pu~\cite{GP}.

In Section~2, we describe the model of nonlinear prices.
In Section~3, we introduce several conditions under which the P\&L of a large trader admits a BSDE representation and the perfect hedging of derivatives is possible. In Section~4, we consider a class of models which admits more explicit computations and verifies the conditions in Section~3.
In Section~5, we consider the hedging of European options and discuss how the model captures illiquidity phenomena.

\section{Model of permanent market impact}
We assume zero risk-free rates.
Let $T>0$ be the end of an accounting period.
Each agent evaluates her utility based on her wealth at $T$.
Consider a security whose value at $T$ is exogenously determined.
We denote the value by $S$  and regard it as an $\mathcal{F}_T$ measurable random variable defined on a filtered probability space $(\Omega, \mathcal{F}, P, \{\mathcal{F}_t\})$ satisfying the usual conditions.
The price of this security at $T$ is trivially $S$, but the price at  $t < T$ should be $\mathcal{F}_t$ measurable and will be endogenously determined by a utility-based mechanism.
There are two agents in our model: A Large trader and a Market.
The Market quotes a price for each volume of the security 
where we have a limit order book in mind.
She can be risk-averse and so her quotes can be nonlinear in volume and depend on her inventory of this security.
The Large trader refers to the quotes and makes a decision. She cannot avoid affecting the quotes by her trading due to the inventory consideration of the Market, and seeks an optimal strategy under this endogenous market impact.

As the pricing rule of the Market, our model adopts the utility indifference principle.
As the utility evaluation of the Market, we consider a family
$\{(\Pi_\tau, \mathcal{D}_\tau)\}_\tau$ of functionals
$\Pi_\tau : \mathcal{D}_T \to \mathcal{D}_\tau$ with the following
properties, where $\tau$ is a
$[0,T]$-valued stopping time and $\mathcal{D}_\tau$ is a linear space of
$\mathcal{F}_\tau$-measurable random variables : For any $X, Y \in \mathcal{D}_T$,
\begin{enumerate}
 \item $\Pi_\tau(0) = 0$,
 \item $\Pi_\tau(X+Y) = \Pi_\tau(X) + Y$ if $Y \in \mathcal{D}_\tau$,
 \item $\Pi_\tau(\lambda X + (1-\lambda)Y) \geq 0 $
       for all $\lambda \in [0,1]$ if 
       $\Pi_\tau(X) \geq 0$ and $\Pi_\tau(Y)\geq 0$,
 \item  $\Pi_\tau(X) \geq \Pi_\tau(Y)$ if there exists 
	$\sigma \geq \tau$ such that
	$\Pi_\sigma(X) \geq \Pi_\sigma(Y)$.
\end{enumerate}
Comments on this axiomatic approach follow in order:
\begin{enumerate}
 \item 
The simplest example is
\begin{equation}\label{rn}
 \Pi_t(X) = E[X|\mathcal{F}_t]
\end{equation}
       with $ \mathcal{D}_t = L^p(\Omega, \mathcal{F}_t,P)$
       with $p \geq 1$. When $p=2$,
 this evaluation can be interpreted as the orthogonal projection of future cash-flows.
 \item
      A more interesting example is an exponential utility :
\begin{equation}\label{exp}
 \Pi_t(X) = -\frac{1}{\gamma} \log E[ \exp\{-\gamma X\} | \mathcal{F}_t] 
\end{equation}
with $  \mathcal{D}_t = L^\infty(\Omega, \mathcal{F}_t,P)$,
      where $\gamma > 0$ is a parameter of risk-aversion.
      By letting $\gamma \to 0$, we recover the previous example.
      By letting $\gamma \to \infty$, we have
      \begin{equation*}
       \Pi_t(X) = \inf\left\{E^Q[X|\mathcal{F}_t] ; Q \sim P, Q=P \text{
		       on } \mathcal{F}_t\right\}.
      \end{equation*}
      which essentially represents the infimum value of $X$ under the conditional
      probability given $\mathcal{F}_t$. By Kupper and
      Schachermayer~\cite{KS},
      no other utility of von
Neumann-Morgenstern type is equivalent to an evaluation
      satisfying the four axioms.   
 \item More generally, $\Pi_t(X) = -\rho_t(X)$ satisfies the four
       axioms, if $\{\rho_t\}$ is a dynamic convex risk measure, see e.g., Barrieu and El
       Karoui~\cite{BE}, Riedel~\cite{Riedel},
       Delbaen~\cite{Delbaen2006},  Delbaen et
       al.~\cite{DPR}, 
       Kl\"oppel and Schweizer~\cite{KlSc}, Cheridito, Delbaen and
       Kupper~\cite{CDK}, Rusczcy\'nsky and Shapiro~\cite{RS}, Detlefsen and Scandolo~\cite{DS}, and Cherny and Madan~\cite{CM09}.  Convex risk measures play an important role for the
       risk managements in
       banks  or insurance companies. 

 \item  When $  \mathcal{D}_t = L^\infty(\Omega, \mathcal{F}_t,P)$,
   under an additional assumption of the so-called Fatou
	     property, $\Pi$ admits a representation
	     \begin{equation*}
	      \Pi_t(X) = \mathrm{ess.inf}\left\{E^Q[X|\mathcal{F}_t] +
					  c_t(Q) ; Q \sim P, Q=P
					  \text{ on } \mathcal{F}_t\right\},
	     \end{equation*}
	     where
	     \begin{equation*}
	      c_t(Q) = \mathrm{ess.sup}\left\{\Pi_t(X) -E^Q[X|\mathcal{F}_t]
					; X \in \mathcal{D}_t\right\}.
	     \end{equation*}
	Based on this representation an agent who uses $\Pi$ as
	her utility evaluation can
	be interpreted as being ambiguity averse in the spirit of the multiple
	priors decision theory of
	Gilboa and Schmeidler~\cite{GS} and the variational preferences of
	Maccheroni et al.~\cite{MMR}, see also Cerreia-Vioglio et al.~\cite{CMMM}.
	In the case of multiple priors
	$c_t(Q)$ can only take the values zero or infinity while
	variational preferences allow for general penalty functions
	$c$. $c_t(Q)$ can be seen as attaching a certain plausibility
	to the model $Q$ at time $t$ with $c_t(Q)=\infty$ meaning that
	the model is fully unreliable and is effectively excluded from
	the analysis. For sufficient and necessary conditions under which such evaluations are time-consistent see for instance \cite{DPR}. Robust expectations of the form above are also known in
	robust statistics, see Huber~\cite{Huber} or the earlier Wald~\cite{Wald}.
	\item In the theory of no-arbitrage pricing, attempts have been made to narrow the no-arbitrage bounds by restricting the set of pricing kernels considered. One of these approaches is the good-deal bounds ansatz introduced in Cochrane and Sa\'{a}-Requejo~\cite{CS00} which corresponds to excluding pricing kernels which induce a too high Sharpe ratio. The intuition is that these deals are ``too good to be true'' and will be eliminated in a competitive market. Using the Hansen-Jagannathan bound it is shown in  \cite{CS00} that this corresponds to only considering pricing kernels which are close to the physical measure in the sense that their variance or in a continuous-time setting their volatility is bounded, see also Bj\"ork and
	Slinko \cite{TBIS}. Hence, the penalty function for a good-deal bound evaluation in a Brownian setting is zero for local martingale measures whose volatility is bounded by a constant $\Lambda>0$ (which depends on the highest possible Sharpe ratio) and infinity else. 
		So if we let $M$ be the set of local equivalent martingale measures and identify each measure $\mathbb{Q} \ll \mathbb{P}$ with a Radon-Nikodym derivative $\frac{d \mathbb{Q}}{d \mathbb{P}} = \mathcal{E}\left((q\cdot W)_T\right)$, with $\mathcal E$ denoting the stochastic exponential,
	we can define $\mathcal{A}^n := \{\mathbb{Q} \ll \mathbb{P} \big| |q|^2 \leq \Lambda\}$. Then the good-deal bound evaluation is given by
	\begin{equation}\label{good_deal}
	\Pi_t(X) = \mathrm{ess.sup}_{\mathbb{Q} \in M \cap \mathcal{A}^n} \mathbb{E}_{\mathbb{Q}}[X|\mathcal{F}_t].
	\end{equation}
	That this evaluation is time-consistent follows for instance from \cite{Delbaen2006}.

\end{enumerate}

We assume $S \in \mathcal{D}_T$ in the sequel. Suppose that the Market is initially endowed with a risky asset which yields a cash-flow at time $T$, represented by $H_{\mathrm{M}} \in \mathcal{D}_T$. If the Market at time $t \in [0, T]$ is holding $z$ units of the security in question and the inventory $H_{\mathrm{M}}$, then her utility is measured  as $\Pi_t(H_{\mathrm{M}} + zS)$.
According to the utility indifference principle,
the Market quotes a selling price for $y$ units of the security by
\begin{equation}\label{pzy}
 \begin{split}
 P_t(z,y) :=& \inf\{ p \in \mathbb{R} ;
  \Pi_t(H_{\mathrm{M}}+zS-yS + p) \geq \Pi_t(H_{\mathrm{M}}+zS) \\
  =& \Pi_t(H_{\mathrm{M}} + zS) - \Pi_t(H_{\mathrm{M}} + (z-y)S).
 \end{split}
\end{equation}
For the equality
we have used the second axiom of $\Pi$ (cash invariance).

Note that in the risk-neutral case (\ref{rn}),
$P_t(z,y) = y E[S|\mathcal{F}_t]$.
In general, 
the price depends on the inventory $z$ of the securities, 
which describes permanent market impact.
In the literature of modeling permanent market impacts,
the absence of price manipulation has been a key issue; see e.g.,
Gu\'eant~\cite{Gueant} and references therein.  Our model does not allow any price manipulation in the sense
that a round-trip cost is always $0$:
\begin{equation*}
 P_t(z,y) + P_t(z-y,-y) = 0.
\end{equation*}
For all $t$ and $z$, $P_t(z,y)$ is a convex function of $y$ with $P_t(z,0) = 0$ by the third axiom of $\Pi$ (concavity).
This implies in particular that
\begin{equation*}
-P_t(z,-y) \leq P_t(z,y) 
\end{equation*}
for any $y$ and $z$, which means that the selling price for an amount is higher than or equal to the buying price for the same amount.
This represents bid-ask spread that is a measure of market liquidity.

Let
$\mathcal{S}_0$ be the set of the simple predictable processes $Y$
with $Y_{0} = 0$.
The Large trader is allowed to take any element $Y \in \mathcal{S}_0$ as her trading strategy.
The price for the $y$ units of the security at time $t$ is
$P_t(-Y_t,y)$. This is because the Market holds $-Y_t$ units of the
security due to the preceding trades with the Large trader. 
Then the profit and loss at time $T$  associated with $Y \in \mathcal{S}_0$  (i.e., the terminal wealth corresponding to the self-financing strategy $Y$)
is given by
\begin{equation*}
 \mathcal{I}(Y) := Y_TS - \sum_{0 \leq t < T} P_t(-Y_t, \Delta Y_t). 
\end{equation*}
Due to (\ref{pzy}), $\mathcal{I}(Y)$ has the form of a nonlinear
stochastic integral studied in Kunita~\cite{Kunita}; see (\ref{iy}) below.
Note that in the risk-neutral case (\ref{rn}), 
\begin{equation*}
 \mathcal{I}(Y) = Y_TS_T - \sum_{0 \leq t < T} S_t \Delta Y_t
  = \int_0^T Y_t \mathrm{d}S_t
\end{equation*}
by integration-by-parts, where
\begin{equation*}
 S_t = E[S|\mathcal{F}_t].
\end{equation*}
In Section~3, we show that $\mathcal{I}(Y)$ admits a representation 
in terms of BSDEs when $\Pi$ is a $g$-expectation,
which enables us to extend the domain $\mathcal{S}_0$ to
a larger set $\mathcal{S}$ of predictable processes.
Now, suppose that the Large trader has an option contract which amounts to pay $-H_\mathrm{L} \in \mathcal{D}_T$ at $T$.
The hedging problem is then to find a unique element $(a, Y) \in \mathbb{R}\times \mathcal{S}$ such that
\begin{equation*}
 -H_\mathrm{L} = a + \mathcal{I}(Y).
\end{equation*}
\section{Hedging in a market with $g$-expectation}
In a continuous-time setting where the filtration is generated by a
Brownian motion it is well known that $\Pi$ satisfying our axioms is essentially equivalent to $\Pi$ being a so called $g$-expectation. $g$-expectations also give a convenient representation of $\mathcal{I}(Y)$.
More precisely, we work under the following condition on the utility
function $(\Pi_t, \mathcal{D}_t)$:
\begin{cond}\label{cond1}
 The filtration $\{\mathcal{F}_t\}$ is the augmentation of the one generated by a standard Brownian motion $W$. 
 Let $g = \{g_t(z)\}: \Omega \times [0,T] \times \mathbb{R} \to
 \mathbb{R}$ be a $\mathcal{P}\otimes \mathcal{B}(\mathbb{R})$
 measurable function, where $\mathcal{P}$ is the progressively measurable $\sigma$ field,
such that 
 $z \mapsto g_t(z)(\omega)$ is a convex function with $g_t(0)(\omega)=0$ for each $(\omega ,t) \in \Omega \times [0,T]$ .
 For each $X \in \mathcal{D}_T$,
\begin{equation*}
 \sup_{0 \leq t \leq T}|\Pi_t(X)| \in \mathcal{D}_T,
\end{equation*}
 and
 there exists a progressively measurable process $Z(X)$ such that
 \begin{equation*}
  E[\int_0^T|Z_t(X)|^2 \mathrm{d}t] < \infty,
 \end{equation*}
 and
\begin{equation}\label{bsdeX}
  X = \Pi_t(X) + \int_t^T g_s(Z_s(X))\mathrm{d}s - \int_t^T Z_s(X) \mathrm{d}W_s,
\end{equation}
 for all $t \geq 0$.
\end{cond}
\begin{exa}
 \upshape
 Let $\mathcal{D}_t = L^2(\Omega,\mathcal{F}_t,P)$ and $G$ be
a progressively measurable process such that
\begin{equation*}
 E\left[\exp\left\{\frac{1}{2}\int_0^T G_t^2 \mathrm{d}t\right\}\right] < \infty.
\end{equation*}
  If $\Pi(X)$ follows (\ref{bsdeX}) with $g_s(z) = G_sz$, then
 \begin{equation}\label{lin}
  X = \Pi_t(X) - \int_t^T Z_s(X) \mathrm{d}W^G_s
 \end{equation}
 and $W^G$ is a standard Brownian motion under $Q$,  where
 \begin{equation*}
  W^G_t = W_t - \int_0^t G_s \mathrm{d}s, \ \
   \frac{\mathrm{d}Q}{\mathrm{d}P}= \exp \left\{
\int_0^T G_t \mathrm{d}W_t - \frac{1}{2}\int_0^T G_t^2 \mathrm{d}t
					 \right\}.
 \end{equation*}
 Therefore,
 \begin{equation*}
  \Pi_t(X) = E^Q[X|\mathcal{F}_t].
 \end{equation*}
 Conversely, if $\Pi(X)$ is defined as the conditional expectation w.r.t. $Q$, then by the martingale representation theorem, there exists $Z(X)$ such that (\ref{lin}) holds, which is equivalent to (\ref{bsdeX}) with $g_s(z) = G_sz$.
\end{exa}
\begin{exa} \label{exa2}
 \upshape
 Let $\gamma > 0$ and 
 $\mathcal{D}_t = \{ X \in L^0(\Omega,\mathcal{F}_t,P) ;  E[\exp\{a|X|\}] < \infty \text{ for all } a > 0\}$, which is an Orlicz space. 
 If $\Pi(X)$ follows (\ref{bsdeX}) with $g_s(z) = \gamma z^2/2$, then
 \begin{equation} \label{expM}
  \mathrm{d}M_t = \gamma M_t Z_t(X) \mathrm{d}W_t,
 \end{equation}
 where $M_t = \exp\{-\gamma \Pi_t(X)\}$. This implies
 \begin{equation*}
  E[\exp\{-\gamma X\}|\mathcal{F}_t] = \exp\{-\gamma \Pi_t(X)\},
 \end{equation*}
 which is equivalent to (\ref{exp}).
 Conversely if $\Pi(X)$ is given by (\ref{exp}), then again by the martingale representation theorem, there exists $Z(X)$ such that (\ref{expM}) holds, which implies (\ref{bsdeX}).
\end{exa}
\begin{exa}
	\upshape
In the good-deal bound example,	suppose that we have a $d$-dimensional Brownian motion $W$ generating the economic noise and that the dynamics of the stock process is given by
	\begin{equation*} 
	\frac{dS^i_t}{S_t^i}=\mu^i dt+\sigma^i dW_t
	,\,\,\,\,\,\,\,\,i=1,\ldots,k. \end{equation*}
	We further suppose that the interest rate of the bond is zero. Let $A:=(\sigma^1,\ldots,\sigma^k)$ and $b:=-\mu^\intercal=-(\mu^1,\ldots,\mu^k)^\intercal$.
	Let $P_B(0)$ be the projection of $0$ onto the
	set $B:=\{x|Ax=b\}$ in the Euclidean $|\cdot|$ norm, and define $P_{\text{Kernel}(A)}(z)$ accordingly as the projection of $z$ in the $|\cdot|$ norm onto the space given by the kernel of the matrix $A$.
	One can prove (see \cite{KLL15}) that the evaluation $\Pi$ in (\ref{good_deal}) is given by a $g$-expectation following (\ref{bsdeX}) with driver function 
	\begin{align*}
	g(t,z)=-\sqrt{\Lambda-|P_B(0)|^2}\Big|P_{\text{Kernel}(A)}(z)
	\Big|+zP_B(0).
	\end{align*}
	This concludes our examples.
\end{exa}
We remark that if
$g_t(z)$ is Lipschitz in $z$ uniformly in $(\omega,t ) \in \Omega \times [0,T]$, then there
exists a unique solution $(\Pi(X),Z(X))$ to (\ref{bsdeX}) for
$X \in L^2(\Omega, \mathcal{F}_T,P)$ and $\Pi_t(X) \in
L^2(\Omega,\mathcal{F}_t,P)$ and the four axioms of $\Pi$ are automatically
satisfied.
As mentioned above it is worthwhile  to note that when the filtration is generated by a standard Brownian
motion,
under additional compactness or domination assumptions, every
evaluation $\Pi$ satisfying our axioms corresponds to a $g$-expectation
in the sense that there exists $g$ such that
$\Pi$ satisfies (\ref{bsdeX}).
For these and other related results, see Jiang~\cite{J08}, Barrieu and El
Karoui~\cite{BE2}, Coquet et al.~\cite{CHMP},
Briand and Hu~\cite{BH1,BH2}, Hu et al.~\cite{HMPY} and the references therein.\\

\noindent
Let
$\Pi^y = \Pi(H_\mathrm{M} -yS)$ and
$Z^y = Z(H_\mathrm{M}-yS)$ for $y \in \mathbb{R}$.
We pose the following technical condition:
\begin{cond}\label{cond2}
 There exist $\Omega_0 \in \mathcal{F}$ with $P(\Omega_0) = 1$ and 
 a $\mathcal{P}\otimes \mathcal{B}(\mathbb{R})$ measurable function
 $$Z : \Omega \times [0,T] \times \mathbb{R}\to \mathbb{R}$$ 
  such that
       $Z(\omega, t,y) = Z^y_t(\omega)$ for all
       $(\omega, t , y) \in \Omega_0 \times [0,T] \times \mathbb{R}$.  
\end{cond}
We will see this condition is always satisfied for Markov models
considered in Section~4. Even in non-Markov cases,
it follows for instance from Ankirchner et al.~\cite{AID}
that if
$g_t(z)$ and its derivative in $z$ are globally Lipschitz and
$H_M$ and $S$ are bounded, then
$Z^y_t(\omega)$ is continuous in $t$ and
differentiable in $y$ for almost all $\omega$,
 which in particular verifies Condition~\ref{cond2}.

\begin{lem} \label{repni} 
 Under Conditions \ref{cond1} and \ref{cond2},
 \begin{equation*}
  \mathcal{I}(Y)
   = H_\mathrm{M} - \Pi_0(H_\mathrm{M}) - \int_0^T g_t(Z^Y_t)\mathrm{d}t
   + \int_0^T Z^Y_t\mathrm{d}W_t
 \end{equation*}
 for $Y \in \mathcal{S}_0$,
 where $Z^Y_t(\omega) = Z(\omega, t, Y_t(\omega))$.
\end{lem}
{\it Proof : } Denote the discontinuity points of $Y \in \mathcal{S}_0$ by
\begin{equation*}
 0 \leq \tau_1 < \tau_2 < \cdots.
\end{equation*}
Let $n$ be the number of the discontinuity points, $\tau_0 = 0$ and
$\tau_k = T$ for  $k \geq n+1$.
By definition,
\begin{equation}\label{iy}
 \begin{split}
  \mathcal{I}(Y) = &
  Y_TS - \sum_{0 \leq t < T}(\Pi_t(H_\mathrm{M} -Y_tS ) - \Pi_t(H_\mathrm{M}-Y_{t+}S)) \\
  =& Y_TS - \sum_{j=1}^n
  (\Pi_{\tau_j}(H_\mathrm{M} -Y_{\tau_j}S ) - \Pi_{\tau_j}(H_\mathrm{M}-Y_{\tau_{j+1}}S)) \\
  =& H_\mathrm{M} - \Pi_0(H_\mathrm{M}) - 
  \sum_{j=0}^n (\Pi_{\tau_{j+1}}
  (H_\mathrm{M} -Y_{\tau_{j+1}}S ) - \Pi_{\tau_j}(H_\mathrm{M}-Y_{\tau_{j+1}}S)).
 \end{split}
\end{equation}
Here we have used that $\Pi_T(H_\mathrm{M}-Y_TS) = H_\mathrm{M}-Y_TS$ and
$Y_0 = 0$. Again by definition,
\begin{equation*}
 \Pi_{\tau_{j+1}}(H_\mathrm{M}-yS) - \Pi_{\tau_j}(H_\mathrm{M}-yS)
  = \int_{\tau_j}^{\tau_{j+1}} g_s(Z^y_s)\mathrm{d}s
  - \int_{\tau_j}^{\tau_{j+1}}Z^y_s\mathrm{d}W_s.
\end{equation*}
Since $Y$ is a simple predictable process,
$Y_{\tau_{j+1}}$ is $\mathcal{F}_{\tau_j}$ measurable and so,
we can substitute $y = Y_{\tau_{j+1}}$ to obtain
\begin{equation*}
 \mathcal{I}(Y)
  =  H_\mathrm{M} - \Pi_0(H_\mathrm{M}) - 
  \sum_{j=0}^n \left\{\int_{\tau_j}^{\tau_{j+1}} g_s(Z^{Y}_s)\mathrm{d}s
  - \int_{\tau_j}^{\tau_{j+1}}Z^Y_s\mathrm{d}W_s \right\},
\end{equation*}
which implies the result. \hfill////\\

\noindent
By this lemma, we naturally extend the domain of $\mathcal{I}(Y)$ to
\begin{equation*}
 \mathcal{S} :=
  \left\{ Y: \Omega\times [0,T] \to \mathbb{R} ;
  \text{ predictable with }
   \int_0^T |Z^Y_t|^2 \mathrm{d}t < \infty 
  \right\}.
\end{equation*}
Now we are ready to give the main result of the paper in an abstract framework.
\begin{cond}\label{cond3}
  There exist $\Omega_0 \in \mathcal{F}$ with $P(\Omega_0) = 1$ and 
 a $\mathcal{P}\otimes \mathcal{B}(\mathbb{R})$ measurable function
 $$Z^- : \Omega \times [0,T] \times \mathbb{R}\to \mathbb{R}$$ 
  such that
       $Z(\omega, t,Z^-(\omega,t,z)) = z$ for all
       $(\omega, t , z) \in \Omega_0 \times [0,T] \times \mathbb{R}$.  
\end{cond}
 \begin{thm}\label{thm1}
  Under Conditions~\ref{cond1}, \ref{cond2} and \ref{cond3}, for any $H_\mathrm{L} \in \mathcal{D}_T$,
  we have
  \begin{equation*}
   - H_\mathrm{L} = 
    \Pi_0(H_\mathrm{M}) - \Pi_0(H_\mathrm{M} + H_\mathrm{L})+ \mathcal{I}(Y^\ast),
  \end{equation*}
  where $Y^\ast$ is defined by $Y^\ast_t(\omega) = Z^-(\omega,t,Z_t(H_\mathrm{M} + H_\mathrm{L})(\omega))$.
 \end{thm}
 {\it Proof : }
 By Condition~\ref{cond1}, there exists
 $Z^\ast := Z(H_\mathrm{M} + H_\mathrm{L})$ such that
   \begin{equation*}
  H_\mathrm{M} + H_\mathrm{L}
   = \Pi_0(H_\mathrm{M} + H_\mathrm{L})
   + \int_0^T g_s(Z^\ast_s)\mathrm{d}s  - \int_0^T Z^\ast_s \mathrm{d}W_s.
   \end{equation*}
  Define $Y^\ast$ as $Y^\ast_t(\omega) = Z^-(\omega,t,Z^\ast_t(\omega))$.
 Then,
 \begin{equation*}
  Z^{Y^\ast}_t(\omega) =
   Z(\omega,t,Y^\ast_t(\omega)) = Z^\ast_t(\omega).
 \end{equation*}
 Therefore, by Lemma~\ref{repni}
 \begin{equation*}
  \mathcal{I}(Y^\ast) =
   H_\mathrm{M} - \Pi_0(H_\mathrm{M}) - \int_0^Tg_t(Z^\ast_t)\mathrm{d}t
   + \int_0^T Z^\ast_t\mathrm{d}W_t,
 \end{equation*}
 which implies the result. \hfill////\\

 \noindent
 This theorem means that any option payoff $-H_\mathrm{L}$ can be perfectly replicated by a self-financing dynamic trading strategy of the security with initial capital
 $$\Pi_0(H_\mathrm{M}) - \Pi_0(H_\mathrm{M} + H_\mathrm{L}).$$
 This is an increasing and convex function of $-H_\mathrm{L}$, which reflects
 a diversification effect of risk. In Section~4, we study even more tractable models and see that Conditions~\ref{cond1}, \ref{cond2} and \ref{cond3} are satisfied under reasonable assumptions.
 
\section{Markov markets}
Here we verify Conditions~\ref{cond2} and \ref{cond3} and
characterize the hedging strategy in terms of solutions of semi-linear
PDEs under Markov models. More precisely, in addition to
Condition~\ref{cond1}, we suppose $g_t(z) = g(z,t)$, 
$S= s(F_T)$ and $H_\mathrm{M}= h_\mathrm{M}(F_T)$, 
where $g : \mathbb{R}\times [0,T] \to \mathbb{R}$,
$s : \mathbb{R} \to \mathbb{R}$ and
$h_\mathrm{M} : \mathbb{R} \to \mathbb{R}$ are Borel functions and 
$F$ is the solution of the SDE
\begin{equation*}
\mathrm{d}F_t = \mu(F_t,t)\mathrm{d}t + \sigma(F_t,t) \mathrm{d}W_t,
\end{equation*}
where $\mu: \mathbb{R} \times [0,T] \to \mathbb{R}$ and $\sigma:
 \mathbb{R} \times [0,T] \to \mathbb{R}^+$ are  Lipschitz functions
 in the following sense: there exists $L > 0$ such that
 \begin{enumerate}
  \item  $|\mu(x,t)-\mu(y,t)| +
	|\sigma(x,t)-\sigma(y,t)| \leq L|x-y|$ and
  \item $|\mu(x,t)| + |\sigma (x,t)| \leq L(1 + |x|)$
 \end{enumerate}
 for all $x,y\in
	 \mathbb{R}$ and $t \in [0,T]$.
The Markov process $F$ should be understood as an economic factor.
As in Section~3, the filtration $\{\mathcal{F}_t\}$ is supposed to be
generated by the standard Brownian motion $W$.
Let $p : \mathbb{R} \times [0,T] \times \mathbb{R} \to \mathbb{R}$ be a $C^{2,1,0}$
solution of the PDE
 \begin{equation}\label{pde}
  \partial_tp(x,t,y) + \mu(x,t) \partial_xp(x,t,y) + \frac{1}{2}\sigma^2(x,t)\partial_x^2p(x,t,y) =
   g(-{ \sigma(x,t)}\partial_xp(x,t,y), t)
 \end{equation}
 on $\mathbb{R}\times (0,T) \times \mathbb{R}$
 with $p(x,T,y) = h_\mathrm{M}(x)-y s(x)$ for all $(x,y) \in \mathbb{R}^2$.
 Here its existence is assumed.
 Then, it is well-known, and easy to check, that
 $(\Pi^y,Z^y)$ defined by
 \begin{equation*}
  \Pi^y_t = p(F_t,t,y), \ \
   Z^y_t = -\sigma(F_t,t)\partial_xp(F_t,t,y)
 \end{equation*}
 is a solution of the BSDE (\ref{bsdeX}) with $X = H_\mathrm{M}-yS$ for each $y \in \mathbb{R}$.
 In the following two subsections, we separately deal with the cases that the driver $g$ is Lipschitz and that $g$ is a quadratic function, or equivalently that $\Pi$ is an exponential utility.

 \subsection{Lipschitz drivers}
   \begin{thm}\label{theorem2}
Let $\mathcal{D}_t = L^2(\Omega,\mathcal{F}_t,P)$ for $t \in [0,T]$ and assume 
    \begin{enumerate}
     \item $h_\mathrm{M}$ and $s$ are in $C^1(\mathbb{R})$ with $s^\prime \geq 0$, $s^\prime(F_T) \in \mathcal{D}_T$,
     \item
	  $\mu$, $\sigma$, and $g$ are in $C^{1,0}(\mathbb{R}\times
	  [0,T])$ and $\partial_zg$ is bounded,
	  \item 
  $p$ is in $C^{3,1,0}(\mathbb{R}\times [0,T] \times \mathbb{R})$ and satisfies
		(\ref{pde}),
  \item  $h_\mathrm{M}^\prime$ is of exponential growth and $\sigma$ and
	 $\mu$ are bounded, or $h_\mathrm{M}^\prime$ is of
	 polynomial growth,
    \end{enumerate}
     and that either one of the following conditions holds,
    \begin{itemize}
     \item [a)] $\inf_{x \in \mathbb{R}}s^\prime(x) > 0$.
     \item [b)] $1/\sigma$ is bounded and for all $t \in [0,T)$, there exists $M \in \mathbb{R}$ such that
	   \begin{equation*}
	    \inf_{x \in [M,\infty)}s^\prime(x) > 0
	   \end{equation*}
	   and 
	   the support of $f + F_T -  F_t$ under $P(\cdot|F_t = f)$ includes
	   $[M,\infty)$ for any $f$ in the support of $F_t$.
	   
\item [c)]  $1/\sigma$ is bounded for all $t \in [0,T)$, there exists $M \in \mathbb{R}$ such that
	   \begin{equation*}
	    \inf_{x \in (-\infty,M]}s^\prime(x) > 0
	   \end{equation*}
      and
      the support of $f + F_T -  F_t$ under $P(\cdot|F_t = f)$ includes
	   $(-\infty,M]$ for any $f$ in the support of $F_t$.
\item [d)] $\sigma(x,t) = \sigma(t)$ is independent of $x$, $\mu$ is
      bounded, and for all $t \in [0,T)$,
      there exists an interval  $[a,b]$ such that
      $b-a >
      2(\|\mu\|_\infty + \|\partial_x\sigma\|_\infty + \|\partial_zg\|_\infty)T$,
      \begin{equation*}
          \inf_{x \in [a,b]}s^\prime(x) > 0,
      \end{equation*}
and       the support of $f+F_T -  F_t$ under $P(\cdot|F_t = f)$ includes
	   $[a,b]$ for any $f$ in the support of $F_t$.
     
    \end{itemize}
    Then, Conditions~\ref{cond2} and \ref{cond3} hold with
    \begin{equation*}
     Z(\omega,t,y) = -\sigma(F_t(\omega),t)\partial_xp(F_t(\omega),t,y), \ \
       Z^-(\omega,t,z) = \inf\{ y \in \mathbb{R} ; Z(\omega,t,y) \geq z\}.
    \end{equation*}
In particular, for any $H_\mathrm{L} \in \mathcal{D}_T$,
    \begin{equation*}
     -H_\mathrm{L} = p(F_0,0,0) - \Pi^\ast_0 + \mathcal{I}(Y^\ast),
      \ \
      Y^\ast_t(\omega) = Z^-(\omega,t,Z^\ast_t), 
    \end{equation*}
    where $(\Pi^\ast,Z^\ast)$ is the unique solution of the BSDE
    \begin{equation}\label{eq:opt}
     H_\mathrm{L} + h_\mathrm{M}(F_T)
      = \Pi^\ast_t + \int_t^T g(Z^\ast_s,s)\mathrm{d}s - \int_t^T Z^\ast_s\mathrm{d}W_s.
    \end{equation}
   \end{thm}
\begin{rmk}
 
   The conditions on $F$ in the cases b)-d) are satisfied if
   the increments $F_T-F_t$ have full support in $\mathbb{R}$ under
   every initial condition $F_t = f$.
\end{rmk}
\begin{rmk}
 Theorem~\ref{theorem2} remains true if  $s^\prime$ is
 replaced with $-s^\prime$ in the assumptions.
\end{rmk}  
   \noindent
   {\it Proof : }
   The unique existence of
   $(\Pi^\ast,Z^\ast)$ follows from the fact that $g$ is Lipschitz as mentioned before.
   Condition~\ref{cond2} follows from the PDE (\ref{pde}) and the continuity of $\partial_xp$.
   To verify Condition~3, we are going to show
   \begin{equation}\label{surj}
    \lim_{y \to \pm \infty} Z^y_t = \pm \infty.
   \end{equation}
Let  $q(x,t,y) = -\partial_xp(x,t,y)$ and differentiate the PDE (\ref{pde}) to obtain,
\begin{equation*}
    \begin{split}
     \partial_tq(x,t,y) &+ \mu(x,t)\partial_xq(x,t,y) + \frac{1}{2}
     \sigma^2(x,t)\partial_x^2 q(x,t,y)
     \\  = & -{(\partial_x\mu(x,t) +
     \partial_zg(\sigma(x,t)q(x,t,y),t)
     \partial_x\sigma(x,t))}q(x,t,y) \\ &-\left(\partial_zg({\sigma(x,t)}q(x,t,y),t)+
     \partial_x\sigma(x,t)\right){\sigma(x,t)}\partial_xq(x,t,y).
     \end{split}
   \end{equation*}
Applying It$\hat{\text{o}}$'s formula to $V^y_t=q(F_t,t,y)\bigg(=\sigma^{-1}(F_t,t)Z_t^y\bigg)$,
\begin{equation*}
 \begin{split}
  V^y_T =& V^y_t + \int_t^T -\bigg[{\left(\partial_x\mu(F_s,s) +
  \partial_zg(Z^y_s,s)\partial_x\sigma(F_s,s)\right)}V^y_s \\ 
  &+\left(\partial_zg({Z^y_s},s)+\partial_x\sigma(F_s,s)\right){ \sigma(F_t,t)}\hat{Z}^y_s
  \bigg]\mathrm{d}s 
  - \int_t^T\sigma(F_t,t) \hat{Z}^y_s \mathrm{d}W_s \\
  =& V^y_t + \int_t^T -{\left(\partial_x\mu(F_s,s) +
  \partial_zg(Z^y_s,s)\partial_x\sigma(F_s,s)\right)}V^y_s\mathrm{d}s- \int_t^T\sigma(F_t,t) \hat{Z}^y_s \mathrm{d}W^Q_s,
 \end{split}
\end{equation*}
where
\begin{equation*}
 \hat{Z}^y_t = -\partial_x q(F_t,t,y), \ \
 W^Q_t = W_t + \int_0^t ({\partial_zg(Z^y_s,s)}+\partial_x\sigma(F_s,s))\mathrm{d}s.
\end{equation*}
Define a probability measure $Q$ (which depends on $y$) by
\begin{equation*}
 \frac{\mathrm{d}Q}{\mathrm{d}P} = \exp\left\{-
\int_0^T ({\partial_zg(Z^y_s,s)}+\partial_x\sigma(F_t,t))\mathrm{d}W_t - \frac{1}{2} \int_0^T({\partial_zg(Z^y_s,s)}+\partial_x\sigma(F_t,t))^2\mathrm{d}t
				       \right\}.
\end{equation*}
Then
\begin{equation*}
 \begin{split}
V^y_t =& E^Q\left[
\exp\left\{\int_t^T{(\partial_x\mu(F_s,s) +
  \partial_zg(Z^y_s,s)\partial_x\sigma(F_s,s))}\mathrm{d}s\right\}V^y_T \bigg| \mathcal{F}_t 
	      \right]
 \\ =& - E^Q\left[
     \exp\left\{\int_t^T{(\partial_x\mu(F_s,s) +
  \partial_zg(Z^y_s,s)\partial_x\sigma(F_s,s))}\mathrm{d}s\right\}h_\mathrm{M}^\prime(F_T) \bigg| \mathcal{F}_t 
    \right] \\ &+
yE^Q\left[
     \exp\left\{\int_t^T{(\partial_x\mu(F_s,s) +
  \partial_zg(Z^y_s,s)\partial_x\sigma(F_s,s))}\mathrm{d}s\right\}s^\prime(F_T) \bigg| \mathcal{F}_t
    \right].  
 \end{split}
\end{equation*}
Note that $Q$ depends on $y$. Under $Q$,
\begin{equation*}
 \mathrm{d}F_t = (\mu(F_t,t)   - {\partial_zg(\sigma(F_t,t)q(F_t,t,y),t)}-\partial_x\sigma(F_t,t))\mathrm{d}t + \sigma(F_t,t)\mathrm{d}W^Q_t
\end{equation*}
and in particular, $F$ is Markov. Note that $F$ under every $Q$ has a
different distribution. Since $\|\partial_x\mu\|_\infty + \|\partial_zg\|_\infty\|\partial_x\sigma\|_\infty < \infty$ and $s^\prime \geq 0$,
it is sufficient to show
\begin{equation}\label{eq9} \sup_{y \in \mathbb{R}} E^Q\left[
|h_\mathrm{M}^\prime(F_T)| | F_t = f
\right]< \infty
\end{equation}
and
\begin{equation}\label{eq8}
 \inf_{y \in \mathbb{R}} E^Q\left[
     s^\prime(F_T) | F_t = f
    \right] > 0.
\end{equation}
Let fix $t \in [0,T)$ and $f$ in the support of $F_t$.
Define $\underbar{F}^Q_u$ and $\bar{F}^Q_u$, $u\geq t$ by
\begin{equation*}
\mathrm{d}\underbar{F}^Q_u = (\mu(\underbar{F}^Q_u,u)-K)\mathrm{d}u
 +\sigma(\underbar{F}^Q_u,u) \mathrm{d}W^Q_u, \ \ \underbar{F}^Q_t = f, 
\end{equation*}
\begin{equation*}
\mathrm{d}\bar{F}^Q_u = (\mu(\bar{F}^Q_u,u)+K)\mathrm{d}u
 +\sigma(\bar{F}^Q_u,u) \mathrm{d}W^Q_u, \ \ \bar{F}^Q_t = f,
\end{equation*}
where $K = \|\partial_zg\|_\infty + \|\partial_x\sigma\|_\infty$.
By Proposition~2.18 in Section~5 of \cite{BMSC} we have
$\underbar{F}^Q_u \leq {F}_u \leq \bar{F}^Q_u$ for all $u \in [t,T]$.
To check Equation~(\ref{eq9}) note that if $h_\mathrm{M}^\prime$ grows
at most exponentially and $\mu$ and $\sigma$ are bounded, we have
\begin{align*}
\sup_{y \in \mathbb{R}} & E^Q\left[
|h_\mathrm{M}^\prime(F_T)| | F_t = f
\right]\\
\leq &L\sup_{y \in \mathbb{R}}E^Q\left[
\exp(C\bar{F}^Q_T)|  \bar{F}^Q_t = f
\right] + L\sup_{y \in \mathbb{R}}E^Q\left[
\exp(-C\underbar{F}^Q_T)|  \underbar{F}^Q_t = f
\right]\\
\leq &  \tilde{L} \sup_{y \in \mathbb{R}}E^Q\left[
\exp\left\{C\int_t^T\sigma(\bar{F}^Q_s,s)\mathrm{d}W_s-\frac{C^2}{2}\int_t^T\sigma^2(\bar{F}^Q_s,s)\mathrm{d}s\right\}|  \bar{F}^Q_t = f
 \right]\\
 & + \tilde{L} \sup_{y \in \mathbb{R}}E^Q\left[
\exp\left\{-C\int_t^T\sigma(\underbar{F}^Q_s,s)\mathrm{d}W_s-\frac{C^2}{2}\int_t^T\sigma^2(\underbar{F}^Q_s,s)\mathrm{d}s\right\}|  \underbar{F}^Q_t = f
\right]  <\infty
\end{align*}
for some constants $L, C, \tilde{L} > 0$,
where the last inequality holds by Novikov's criterion. A similar
argument holds as well for the case that
 $h_\mathrm{M}^\prime$ is of polynomial growth without the boundedness
 of $\mu$  and $\sigma$, where we use that
\begin{equation*}
 E^Q[|\bar{F}^Q_T|^m | \bar{F}^Q_t = f] < \infty, \ \
   E^Q[|\underbar{F}^Q_T|^m | \underbar{F}^Q_t = f] < \infty
\end{equation*}
for any $m \in \mathbb{N}$.
Note that the left hand sides do not depend on $Q$ and so, also not on
$y$.
To check (\ref{eq8}), we consider the four cases in order.

 \textbf{Case a)}:
 In this case (\ref{eq8}) is clear.

 \textbf{Case b)}:
 Suppose that $s'(x) \geq \epsilon > 0$ for all $x \in [M,\infty)$. 
Clearly, $\underbar{F}^Q$ under every $Q$ has the same distribution and
the same holds for $\bar{F}^Q$.
Hence,
\begin{equation*}
E^Q[s^\prime(F_T)| F_t = f]
 \geq
E^Q[s^\prime(F_T)1_{[M,\infty)}(\underbar{F}^Q_{T})| F_t = f]
\geq \epsilon
Q(\underbar{F}^Q_T \geq M)> 0, 
\end{equation*}
where the last strict inequality holds as $\underbar{F}^Q_T$ has the same
distribution under $P'$ given by
\begin{equation*}
 \frac{\mathrm{d} P'}{\mathrm{d}Q} = \exp\left\{
					  K\int_t^T \frac{\mathrm{d}W^Q_u}{\sigma(\underbar{F}^Q_u,u)}
					  -
					  \frac{K^2}{2}
					  \int_t^T \frac{\mathrm{d}u}{\sigma(\underbar{F}^Q_u,u)^2}
					 \right\}
\end{equation*}
 as $F_T$ under $P(\cdot | F_t = f)$.
 The probability measure $P'$ is well-defined because of the boundedness
 assumption on $1/\sigma$.
 The rest follows from Theorem~\ref{thm1}.

 \textbf{Case c)}: Is treated
 similarly to Case~b).

\textbf{Case d)}: Define $\hat{F}^Q_u$, $u \geq t$ by
\begin{equation*}
\mathrm{d}\hat{F}^Q_u= \mu(\hat{F}^Q_u,u)\mathrm{d}u
 +\sigma(\hat{F}^Q_u,u) \mathrm{d}W^Q_u= \mu(\hat{F}^Q_u,u)\mathrm{d}u
 +\sigma(u) \mathrm{d}W^Q_u, \ \
 \hat{F}^Q_t = f.
\end{equation*}
Clearly, $\hat{F}^Q_T$ under every $Q$ has the same distribution as $F$
under $P(\cdot | F_t = f)$. As $\sigma$ does not depend on $x$,
\begin{equation*}
 \|\hat{F}^Q_T-F_T\|_\infty = 
\|\hat{F}^Q_T-\hat{F}^Q_t - (F_T-F_t)\|_\infty \leq (K+\|\mu\|_\infty)(T-t).
\end{equation*}
Put
\begin{equation*}
  B = \left\{
       x \in \mathbb{R}; \left|x -\frac{a+b}{2}\right| < \frac{b-a}{2}-(K+\|\mu\|_\infty)T
      \right\}.
\end{equation*}
Then,
\begin{equation*}
 \begin{split}
 E^Q[s^\prime(F_T)| F_t = f]
  & \geq
  E^Q[s^\prime(F_T)1_B(F_t  + \hat{F}^Q_{T}-\hat{F}^Q_t)| F_t = f]
  \\ &\geq \epsilon
Q(f+\hat{F}^Q_T-\hat{F}^Q_t\in B)=\epsilon P(f+F_T-F_t\in B | F_t = f) > 0, 
   \end{split}
\end{equation*}
where the second inequality holds since necessarily $f+F_T-F_t\in [a,b]$ if $f+\hat{F}^Q_T-\hat{F}^Q_t \in B$.
 The rest follows from Theorem~\ref{thm1}. \hfill////\\

 \subsection{Exponential utilities}
 \begin{thm}
 Let $\mathcal{D}_t = \{ X \in L^0(\Omega,\mathcal{F}_t,P) ;  E[\exp\{a|X|\}] < \infty \text{ for all } a > 0\}$, $\mu(x,t) = b(t)$, $\sigma(x,t)=\sigma(t)$, and
 $g(z,t) = \beta(t) z + \gamma z^2/2$, where
 $b \in L^2([0,T])$, $\beta$ satisfies $\mathbb{E}\left[\exp\left\{
 \frac{1}{2}\int_0^T \beta(t)^2\mathrm{d}t
 \right\} \right] < \infty $ and $\gamma > 0$.
  If $s$ and $h_\mathrm{M}$ are of linear growth and
  $s$ is strictly monotone on $\mathbb{R}$,
   then Conditions~\ref{cond1}, \ref{cond2} and \ref{cond3} hold.
 \end{thm}
 {\it Proof : }  Extending Example~\ref{exa2}, we have
 \begin{equation*}
  \Pi_t(X) =
   -\frac{1}{\gamma} \log
   E^Q[\exp\{-\gamma X\}|\mathcal{F}_t],
   \ \ \frac{\mathrm{d}Q}{\mathrm{d}P} = \exp\left\{
\int_0^T \beta(t)\mathrm{d}W_t - \frac{1}{2}\int_0^T \beta(t)^2\mathrm{d}t
					     \right\}
 \end{equation*}
 for $X \in \mathcal{D}_T$. In particular, Condition~\ref{cond1} holds and
 \begin{equation*}
  \Pi^y_t =
   -\frac{1}{\gamma}\log E^Q[\exp\{-\gamma(h_\mathrm{M}(F_T)-ys(F_T))\}|\mathcal{F}_t]
 \end{equation*}
with $F_T = \hat{\sigma}_{0,T}W^Q_T + B(T)$, where $W^Q$ is a standard Brownian motion under $Q$ and
 \begin{equation*}
B(t) = \int_0^t (b(s) + \beta(s))\mathrm{d}s \quad \text{and} \quad \hat{\sigma}_{t,T}=\sqrt{\int_t^T\sigma^2(s)\mathrm{d}s}.
 \end{equation*}
 By a straightforward computation, we see,
 \begin{equation*}
  \begin{split}
 & p(x,t,y) \\ &=
   -\frac{1}{\gamma} \log
   \int
   \exp\left\{ -\gamma (h_\mathrm{M}(u) - ys(u))-
\frac{(u-x+B(t)-B(T))^2}{2\hat{\sigma}^2_{t,T}}
			  \right\}
   \frac{\mathrm{d}u}{  \sqrt{2\pi \hat{\sigma}^2_{t,T}}}
  \end{split}
 \end{equation*}
 and so,
 \begin{equation*}
  \begin{split}
   & -\partial_xp(x,t,y) \gamma \exp(-\gamma p(x,t,y))
   \\ &=
   \int    \frac{(u-x+B(t)-B(T)) }{  \sqrt{2\pi }\hat{\sigma}^3_{t,T}}
   \exp\left\{ -\gamma (h_\mathrm{M}(u) - ys(u))-
\frac{(u-x+B(t)-B(T))^2}{2\hat{\sigma}^2_{t,T}}
			  \right\}\mathrm{d}u.
  \end{split}
 \end{equation*}
 Therefore $Z^y = -\partial_xp(F_t,t,y)$ is continuous in $y$ and in particular,
 Condition~\ref{cond2} holds.
 Denote by $(l,r)$ the interval $s(\mathbb{R})$.
 Fix $t \in [0,T)$ and define $\varphi : [l,r]\to [-\infty,\infty]$ by
\begin{equation*}
 \varphi(v) = \frac{s^{-1}(v)-x + B(t) -B(T)}{\hat{\sigma}^3_{t,T}}.
\end{equation*}
Further, define a measure $\mu$ on $(l,r)$ by
\begin{equation*}
 \mu(\mathrm{d}v) =
  \exp\left\{
       -\gamma h_\mathrm{M}(s^{-1}(v))
       - \frac{\left(s^{-1}(v)-x + B(t)+B(T)\right)^2}{2\hat{\sigma}^2_{t,T}}
      \right\} s^{-1}(\mathrm{d}v).
\end{equation*}
Then, by applying Lemma~\ref{ess} in Appendix, we have
$\lim_{y \to \pm \infty} |\partial_xp(x,t,y)| = \infty$,
which implies Condition~3. \hfill////\\

The following proposition shows that the strict monotonicity of $s$ is essential
for Condition~\ref{cond3} to hold under exponential utilities.
This is in contrast to the case of Lipschitz drivers.

  \begin{prop}
   Let $\mu = 0$, $\sigma =1$, $g(z,t) = \gamma z^2/2$, $h_\mathrm{M}=0$ and $s(x) = (x -k)_+$, where $k \in \mathbb{R}$. Then, for any $t \in [0,T)$,
 \begin{equation*}
    \lim_{y \to \infty} Z^y_t = \infty, \ \
     \lim_{y \to - \infty} Z^y_t = - \frac{\phi\left(
\frac{k-W_t}{\sqrt{T-t}}
    \right)}{\gamma \sqrt{T-t}\Phi\left(
\frac{k-W_t}{\sqrt{T-t}}
    \right)}, 
 \end{equation*}
   where $\phi$ and $\Phi$ are the standard normal density and distribution functions respectively.
  \end{prop}
  
  \noindent
  {\it Proof: }
  Since
  \begin{equation*}
p(w,t,y) = 
    -\frac{1}{\gamma} \log E[ \exp(y\gamma (W_T-k)_+) | W_t = w],
  \end{equation*}
  we obtain
  \begin{equation*}
   \begin{split}
  & \exp(- \gamma p(w,t,y))
   \\ &=     \exp\left(\frac{T-t}{2}\gamma^2y^2 + \gamma y (w-k)\right)\left(1-\Phi\left(\frac{k-w}{\sqrt{T-t}}-\sqrt{T-t}\gamma y\right)\right) + \Phi\left(
\frac{k-w}{\sqrt{T-t}}
    \right)
   \end{split}
  \end{equation*}
  and so,
   \begin{equation*}
   \begin{split}
   Z^y_t = & - \frac{\partial p}{\partial w}(W_t,t,y) \\
    = &\frac{ y \exp\left(\frac{T-t}{2}\gamma^2y^2 + \gamma y (W_t-k)\right)\left(1-\Phi\left(\frac{k-W_t}{\sqrt{T-t}}-\sqrt{T-t}\gamma y\right)\right)}
    {\exp\left(\frac{T-t}{2}\gamma^2y^2 + \gamma y (W_t-k)\right)\left(1-\Phi\left(\frac{k-W_t}{\sqrt{T-t}}-\sqrt{T-t}\gamma y\right)\right) + \Phi\left(\frac{k-W_t}{\sqrt{T-t}}
    \right)}.   
   \end{split}
   \end{equation*}
   Here we have used the identity
   \begin{equation*}
    \exp\left(\frac{T-t}{2}\gamma^2y^2 + \gamma y (w-k)\right)
     \phi\left(
\frac{k-w}{\sqrt{T-t}}-\sqrt{T-t}\gamma y
		\right)
     = \phi\left(\frac{k-w}{\sqrt{T-t}}
    \right).
   \end{equation*}
   Since $\Phi(-\infty) = 0$, we have
   $\lim_{y \to \infty} Z^y_t = \infty$. Since
   \begin{equation*}
    \lim_{x \to \infty} \frac{x(1-\Phi(x))}{\phi(x)} = 1,
   \end{equation*}
   we have
   \begin{equation*}
    \lim_{y \to -\infty}
     Z^y_t = \lim_{y \to -\infty}
     \frac{y\phi\left(\frac{k-W_t}{\sqrt{T-t}}
    \right)}{\phi\left(\frac{k-W_t}{\sqrt{T-t}}
    \right) + \left(
\frac{k-W_t}{\sqrt{T-t}}-\sqrt{T-t}\gamma y
		\right)\Phi\left(\frac{k-W_t}{\sqrt{T-t}}
	    \right)}
		= - \frac{\phi\left(
\frac{k-W_t}{\sqrt{T-t}}
    \right)}{\gamma \sqrt{T-t}\Phi\left(
\frac{k-W_t}{\sqrt{T-t}}
    \right)}.
   \end{equation*}
   \hfill////

   \section{Explicit computations for European options}
   Here we consider the case $H_L = h_L(S)$ with a Borel function
   $h_L : \mathbb{R} \to \mathbb{R}$ under the Markov framework of the
   previous section.
   This corresponds to the situation where the Large trader has to
   hedge an European option $-h_L(S)$ written on $S$.
   Then,
   the solution $(\Pi^\ast,Z^\ast)$ of the BSDE (\ref{eq:opt}) is given by
   \begin{equation*}
    \Pi^\ast_t = v(F_t,t), \ \ Z^\ast_t =
     -\sigma(F_t,t)\partial_xv(F_t,t),
   \end{equation*}
   where
   \begin{equation*}
    \begin{split}
&\partial_t v(x,t) + \mu(x,t)\partial_xv(x,t) +
 \frac{1}{2}\sigma^2(x,t)\partial_x^2 v(x,t) =
     g(-\sigma(x,t)\partial_xv(x,t),t), \\
     & v(x,T) = h_M(x) + h_L(s(x)).
     \end{split}
   \end{equation*}

   Now, let us consider a specific model to discuss how our
   consideration of market impacts
   affects hedging strategies.
   Let  $F_t = W_t$, $H_M = a S$ with $a \in \mathbb{R}$,
   $S = b + c W_T$ with $b \in \mathbb{R}$, $c > 0$ and $g(z,t) =
   \gamma z^2/2$  with $\gamma \geq 0$. Then,
    when $\gamma > 0$,
   \begin{equation*}
    \begin{split}
    \Pi^y_t &= p(W_t,t,y)
     = -\frac{1}{\gamma}\log
     E[\exp\left\{-\gamma(a-y)(b + c W_T) \right\}|W_t]
     \\ & = (a-y)(b + c W_t) - \frac{T-t}{2}
     \gamma (a-y)^2c^2.
    \end{split}
   \end{equation*}
   This can be also seen from the fact that
   \begin{equation*}
    p(x,t,y) = (a-y)(b+c x) - \frac{T-t}{2}
     \gamma (a-y)^2c^2
   \end{equation*}
   solves
   \begin{equation*}
    \partial_t p(x,t,y) + \frac{1}{2}\partial_x^2 p(x,t,y)
     = \frac{\gamma}{2} |\partial_x p(x,t,y)|^2, \ \
     p(x,T,y) = (a-y)(b+cx).
   \end{equation*}
   (This remains true when $\gamma = 0$ as well.)
   It is then easy to see that
   \begin{equation*}
    Z^y_t = -(a-y)c,\ \ Z^-(\omega,t,z) = a + \frac{z}{c}
   \end{equation*}
   and so, the hedging strategy for $-h_L(S)$ is
   \begin{equation*}
    Y^\ast_t = a -\frac{1}{c} \partial_x v(W_t,t),
   \end{equation*}
   where $v$ is the solution of
   \begin{equation*}
    \partial_t v(x,t) + \frac{1}{2}\partial_x^2v(x,t) =
     \frac{\gamma}{2} |\partial_x v(x,t)|^2, \ \
     v(x,T) = a(b+cx) + h_L(b+cx).
   \end{equation*}
   Note that this is a backward Kardar-Parisi-Zhang equation and the derivative $u =
   \partial_x v$ solves a backward Burgers' equation:
    \begin{equation} \label{Burg}
    \partial_t u(x,t) + \frac{1}{2}\partial_x^2u(x,t) =
     \gamma u(x,t)\partial_x u(x,t), \ \
     u(x,T) = ac + ch_L^\prime(b+cx).
    \end{equation}
    We also have an integral representation; when $\gamma > 0$,
    \begin{equation*}
     \begin{split}
     v(x,t) &= -\frac{1}{\gamma}
      \log E[\exp\left\{-\gamma (a(b+cW_T) + h_L(b+cW_T))\right\} | W_t
      = x]\\
      &=
     -\frac{1}{\gamma}
      \log  \int \exp\left\{ -\gamma (ay + h_L(y))\right\}
      \frac{1}{\sqrt{2\pi c^2(T-t)}}\exp\left\{-\frac{(y-b-cx)^2}{2c^2(T-t)}\right\}\mathrm{d}y.
     \end{split}
    \end{equation*}
    When $\gamma = 0$,
    \begin{equation*}
     v(x,t) = a(b+cx) + E[h_L(b+cW_T)|W_t=x],
    \end{equation*}
    which corresponds to hedging under the Bachelier model.

    There are some cases where we can be more explicit.
    It is known and easily checked that if $u$ is a solution of a Burgers' equation, then
    $u_\lambda(x,t) =  \lambda u(\lambda x, \lambda^2 t)$ is also a solution
    of
    a Burgers' equation.
 Moreover, some non-trivial explicit solutions are available; for example,
    \begin{equation*}
     u(x,t) = 1- \tanh( \gamma x +  \gamma^2 t + \delta)
    \end{equation*}
    with $\delta \in \mathbb{R}$ and $ 1- \tanh( \gamma x +  \gamma^2 T + \delta)$ being the terminal condition.
    
    Suppose $\gamma > 0$, $a = 0$ and the Large trader has to hedge
    a huge amount of put options ($K \in \mathbb{R}$, $\lambda >>1$)
    \begin{equation*}
     2\lambda (K-S)_+ \approx \lambda \left( K-S + \frac{1}{\lambda \gamma}\log
				      \cosh (-\lambda \gamma(K-S))
				     \right) =: -h_L(S).
    \end{equation*}
    Since
    \begin{equation*}
     h^\prime_L(s) = \lambda (1- \tanh(-\lambda \gamma (K-s))),
    \end{equation*}
   the solution $u$ of (\ref{Burg}) is given by
    \begin{equation}\label{Burgsol}
     u(x,t) = \lambda c (1-\tanh(\gamma\lambda c x 
      + \gamma^2\lambda^2 c^2 t  + \delta)),
    \end{equation}
    where $\delta = \lambda \gamma (b-K)- \gamma^2\lambda^2 c^2T$.
    Hence, the hedging strategy is
    \begin{equation}\label{YG}
     Y^\ast_t = -\lambda (1- \tanh(\gamma\lambda (b + c W_t -K)
      - \gamma^2\lambda^2 c^2(T-t))).
    \end{equation}
    It also follows that
    \begin{equation*}
     \begin{split}
     v(x,t) = & \lambda\Biggl\{
		      b + cx - K
-\lambda\gamma c^2(T-t) \\
&- \frac{1}{\lambda \gamma}\log \cosh \left(
								      \lambda
								      \gamma(b
								      +
								      cx
								      - K)-\lambda^2\gamma^2c^2(T-t)\right)
\Biggr\}
    \end{split}
    \end{equation*}
    and so, by Theorem~\ref{thm1}, the replication cost at time $0$ is computed as
    \begin{equation*}
     \begin{split}
   &  p(W_0,0,0) - v(W_0,0) \\& = \lambda\left\{
				     K-S_0 +\lambda\gamma c^2 T+ \frac{1}{\lambda \gamma}\log \cosh \left(
-\lambda \gamma (K-S_0) - \lambda^2 \gamma^2 c^2T
									      \right)
					    \right\}\\
      & \approx 2\lambda(K -S_0+ \lambda \gamma c^2 T)_+,
     \end{split}
    \end{equation*}
    where $S_0 = b + cW_0$. Here nonlinearity in $\lambda$ is clearly seen.

    On the other hand,
    when $\gamma = 0$, we are under the Bachelier model and so,
    the hedging of put options is standard; putting $S_t =
    E[S|\mathcal{F}_t] = b + cW_t$,
    \begin{equation*}
     E[2 \lambda(K-S)_+|\mathcal{F}_t] = 2\lambda\left(
					      (K-S_t)\Phi\left(\frac{K-S_t}{c\sqrt{T-t}}\right) +
		c\sqrt{T-t}			      \phi\left(\frac{K-S_t}{c\sqrt{T-t}}\right)
					     \right)
    \end{equation*}
    and so the hedging strategy is
    \begin{equation}\label{YB}
     Y^\ast_t = -2 \lambda \Phi\left(\frac{K-S_t}{c\sqrt{T-t}}\right)
      = -2\lambda \Phi\left(\frac{K-b-cW_t}{c\sqrt{T-t}}\right).
    \end{equation}
    Both of (\ref{YG}) and (\ref{YB}) are $(-2\lambda,0)$-valued 
    increasing functions of $S_t$.
    The striking difference is in their dependence on $T-t$.
    While the strategy becomes
    flatter as $T-t$ increases under the Bachelier model (\ref{YB}),  $T-t$ is only a location parameter and
    does not change the functional shape under (\ref{YG}).
    The function (\ref{Burgsol}) is interpreted as a shockwave
    propagated from the terminal condition $h^\prime_L$.
    \appendix
    
\section{Convergence of Esscher measures}
\begin{lem}\label{ess}
 Let $\mu$ be a measure on $\mathbb{R}$ with
 \begin{equation*}
\int (1+|x|)e^{yx} \mu(\mathrm{d}x) < \infty
 \end{equation*}   
 for all $y \in \mathbb{R}$.
 Denote
 \begin{equation*}
 l = \inf \mathrm{supp}(\mu) , \ \ 
r=  \sup \mathrm{supp}(\mu), \ \
-\infty \leq l < r \leq \infty.
 \end{equation*} 
 Define the Esscher measure $\mu^y$ by
 \begin{equation*}
  \mu^y(\mathrm{d}x) = \frac{e^{yx}}{m(y)}  \mu(\mathrm{d}x), \ \
   m(y)=  \int e^{yx} \mu(\mathrm{d}x) 
 \end{equation*}
 and let $\mathcal{J}$ be the set of the nondecreasing Borel functions $\varphi : [l,r] \to [-\infty, \infty]$
 with
 \begin{equation*}
  \int (1+|x|) |\varphi(x)| \mu^y(\mathrm{d}x) < \infty 
 \end{equation*}
 for all $y \in \mathbb{R}$.
 \begin{enumerate}
  \item For any $\varphi \in \mathcal{J}$,
	\begin{equation*}
	 y \mapsto \int \varphi(x)\mu^y(\mathrm{d}x)
	\end{equation*}
	is nondecreasing.
  \item If $l > -\infty$, then $\mu^y$ converges weakly to $\delta_l$ as $y \to -\infty$.
  \item For any $\varphi \in \mathcal{J}$ with
	$\lim_{x \to l} \varphi(x)= -\infty$,
	\begin{equation*}
	 \lim_{ y \to -\infty }
	  \int  \varphi(x)\mu^y(\mathrm{d}x) = -\infty.
	\end{equation*}
	  \item If $r <\infty$, then $\mu^y$ converges weakly to $\delta_r$ as $y \to \infty$.
  \item For any $\varphi \in \mathcal{J}$ with
	$\lim_{x \to r} \varphi(x)= \infty$,
	\begin{equation*}
	 \lim_{ y \to \infty }
	  \int  \varphi(x)\mu^y(\mathrm{d}x) = \infty.
	\end{equation*}
 \end{enumerate}
 Here $\delta_l$ and $\delta_r$ are the delta measures of the points $l$ and $r$ respectively.
\end{lem}
{\it Proof: }
1. Note that
\begin{equation*}
 \begin{split}
 &\frac{\mathrm{d}}{\mathrm{d}y} \int \varphi(x) \mu^y(\mathrm{d}y)
 \\& = \frac{1}{m(y)}\int \varphi(x)xe^{yx}\mu(\mathrm{d}x)
  - \frac{1}{m(y)^2} \int \varphi(x) e^{yx}\mu(\mathrm{d}y) \int x e^{yx}\mu(\mathrm{d}x) \\
  &= \int  \varphi(x)x \mu^y(\mathrm{d}x) - \int \varphi(x)\mu^y(\mathrm{d}x) \int x \mu^y(\mathrm{d}x).
 \end{split}
\end{equation*}
The right hand side sequence is nonnegative by the FKG inequality, or just
because this is 
the covariance of comonotone random variables under the probability measure $\mu^y$.\\

\noindent
2.
Denote
\begin{equation*}
 a(y,u) =  \int_{(-\infty,u]} e^{yx}\mu(\mathrm{d}x), \ \
 b(y,u) =  \int_{(u,\infty)} e^{yx}\mu(\mathrm{d}x).
\end{equation*}
Then for any $y < 0$, $u \in (l,r)$ and $\epsilon \in (0,u-l)$, 
\begin{equation*}
 \frac{a(y,u)}{b(y,u)}
  \geq  \frac{a(y,u-\epsilon)}{b(y,u)}
  \geq
  \frac{\int_{(-\infty,u-\epsilon]} e^{y(u-\epsilon)} \mu(\mathrm{d}x)}{\int_{(u,\infty)} e^{yu} \mu(\mathrm{d}x)} = e^{-y\epsilon} \frac{\mu((-\infty,u-\epsilon])}{\mu((u,\infty))}.
\end{equation*}
It follows then that $a(y,u)/b(y,u) \to \infty$ as $y \to - \infty$ for each $u \in (l,r)$. This implies the convergence of the distribution function
\begin{equation}\label{convdel1}
 \mu^y((-\infty,u]) = \frac{a(y,u)}{a(y,u) + b(y,u)} \to 1
\end{equation}
as $y\to -\infty$ for each $u \in (l,r)$.
Now, assume $l> -\infty$. Then, $\mu^y((-\infty,u]) = 0$ for all $u < l$ and so,
 $\mu^y \to \delta_l$ weakly.\\

\noindent
3. Let $\varphi \in \mathcal{J}$ with $\lim_{x\to l} \varphi(x) = -\infty$.
Then for any $n \in \mathbb{N}$, there exists $\delta > 0$ such that
for all $x < l+\delta$, $\varphi(x) < -n$.
Therefore, 
\begin{equation*}
 \int \varphi(x) \mu^y(\mathrm{d}y) \leq
  -n \mu^y((-\infty,l+\delta]) + \int \varphi^+(x)\mu^y(\mathrm{d}y),
\end{equation*}
where $\varphi^+$ is the positive part of $\varphi$.
Since $\varphi^+ \in \mathcal{J}$, the second term is nondecreasing in $y$ as we have already seen. Together with $(\ref{convdel1})$, it implies
\begin{equation*}
\limsup_{y\to -\infty} \int \varphi(x) \mu^y(\mathrm{d}y) \leq -n +
  \int \varphi^+(x) \mu^0(\mathrm{d}y).
\end{equation*}
Since $n$ can be arbitrary we conclude.\\

\noindent
The proofs for 4 and 5 are similar to those for 2 and 3 respectively.
 \hfill////

\end{document}